\documentclass[aps,prb,groupedaddress]{revtex4}
\usepackage{epsf}

\newcommand{\illuseps}[2]{\begin{figure}
\centerline{\epsffile{#1.eps}}
\caption{#2}
\end{figure}}

\newcommand{\tableeps}[2]{\begin{table}
\centerline{\epsfbox{#1.eps}}
\caption{#2}
\end{table}}

\begin{document}

\title{Using bond-length dependent transferable force constants to predict 
vibrational entropies in Au-Cu, Au-Pd, and Cu-Pd alloys}

\author{Eric J. Wu}
\affiliation{Department of Materials Science and Engineering, Massachusetts Institute of Technology, Cambridge, MA 02139, USA}
\author{Gerbrand Ceder}
\affiliation{Department of Materials Science and Engineering, Massachusetts Institute of Technology, Cambridge, MA 02139, USA}
\author{Axel van de Walle}
\affiliation{Department of Materials Science and Engineering, Northwestern University, Evanston, IL 60208, USA}

\begin{abstract}
A model is tested to rapidly evaluate the vibrational properties of alloys 
with site disorder. It is shown that length-dependent transferable force 
constants exist, and can be used to accurately predict the vibrational 
entropy of substitutionally ordered and disordered structures in Au-Cu, 
Au-Pd, and Cu-Pd. For each relevant force constant, a length- dependent 
function is determined and fitted to force constants obtained from 
first-principles pseudopotential calculations. We show that these 
transferable force constants can accurately predict vibrational entropies of 
L1$_{2}$-ordered and disordered phases in Cu$_{3}$Au, Au$_{3}$Pd, 
Pd$_{3}$Au, Cu$_{3}$Pd, and Pd$_{3}$Au. In addition, we calculate the 
vibrational entropy difference between L1$_{2}$-ordered and disordered 
phases of Au$_{3}$Cu and Cu$_{3}$Pt.
\end{abstract}

\maketitle

\section{Introduction}

In the last decade, a clear prescription has emerged to obtain phase 
diagrams of materials from first-principles. When two compounds mix, some 
amount of site disorder occurs. First-principles alloy theory has been 
mostly preoccupied with defining appropriate energy models for such 
partially disordered systems, and obtaining the parameters for them from 
increasingly accurate Density Function Theory (DFT) methods. In particular, 
the cluster expansion approach $^{1}$ has been highly successful, as it 
allows one to parameterize the energy of systems with disorder on a fixed 
framework of sites. Many binary phase diagrams have been calculated in this 
way $^{2 - 6}$. 

Most of these first-principles phase diagram calculations do not include 
vibrational entropy effects. The assumption that vibrational entropy 
differences between phases are small, and can therefore be neglected, has 
been recently examined experimentally $^{7 - 11}$ and theoretically $^{12 - 
18}$. In Al$_{2}$Cu $^{18}$ vibrational entropy contributions were shown to 
be essential in reversing the stability of $\theta $ and $\theta $' phases 
at finite temperatures. In the Al-Sc system, vibrational entropy 
contributions were shown to increase the solubility limits 27-fold $^{16, 
17}$.

For alloys with partial site disorder, vibrational entropy in 
first-principles phase diagrams is computationally intensive to determine. 
In conventional first-principles alloy theory, a cluster expansion has to be 
fit to the ground state energies of a large number of A-B ordered states 
(for a binary alloy). Vibrational effects can be formally included by 
fitting to vibrational free energies, rather than ground state energies 
$^{19}$. This requires that one determines the phonon spectrum for many 
configurations. Unfortunately, this is computationally expensive.

One idea that has been proposed to alleviate this problem is the use of 
force constants that can be transferred between different environments. The 
existence of transferable force constants would greatly reduce the number of 
calculations required to determine vibrational entropies from 
first-principles: currently, almost all of the work in calculating 
vibrational entropies is involved with the numerical determination of force 
constants for every environment in a large number of structures.

The idea of using transferable force constants to reduce computational 
expense is tremendously appealing. Despite this, until recently, there was 
little clear evidence supporting the validity of their use. For instance, 
experimental data in Fe$_{3}$Al shows that force constants can have a clear 
configurational dependence $^{20}$. Strictly speaking, it is not possible to 
rigorously define configuration-independent force constant matrices as the 
form of the force constant matrix depends on the symmetry of the 
configuration $^{21}$. Nevertheless, configuration-independent transferable 
force constants have been used in computations to study segregation in 
Ni$_{3}$Al $^{22}$ and elastic properties in Li-Al $^{23}$. In Li-Al, these 
force constants predicted elastic constants well, but reproduced vibrational 
entropies poorly. Transferable force constants have also been explored in 
oxides $^{24}$ and semiconductors $^{25}$. In these materials, 
transferability is examined in the short-ranged force constants, which are 
obtained after subtracting out an analytical long-ranged dipole-dipole 
contribution. None of these studies examined the use of transferable force 
constants to calculate vibrational entropy. This is a challenging issue, as 
vibrational entropy differences between structures are a fraction of the 
total vibrational entropy and hence, high relative accuracy is required.

Previous attempts at defining transferable force constant matrices $^{23}$ 
defined force constant matrices for a given pair type. Recently, it was 
shown that high-accuracy in predicting the vibrational entropy could be 
achieved by using bond length-dependent transferable force constants $^{12, 
26}$. This simple bond length-dependent force constant model was shown to 
work well in predicting vibrational entropies in Ni$_{3}$Al and Pd$_{3}$V 
$^{12, 13}$. The force constants were found to be transferable between 
different FCC structures at a single composition; their transferability 
between configurations with different atomic ratios or different chemical 
systems was untested. 

In this paper, we evaluate the accuracy of the length-dependent transferable 
force constants approach in calculating vibrational entropies in the Au-Cu, 
Au-Pd, and Cu-Pd systems, and also investigate the transferability of such 
relations \textit{between} different chemical systems. Our results show that in the Au-Cu, 
Au-Pd, and Cu-Pd systems, transferable force constants can be defined that 
depend on length and pair type. We show that these force constants can be 
exchanged between different structures and chemical systems, while still 
retaining good accuracy for calculating the vibrational entropy of 
intermetallic structures.

\section{Methodology}

A supercell method was used to calculate force constants $^{12,26 - 29}$. 
This method consists of creating a series of supercells with well-chosen 
displacements, calculating the resulting forces on atoms, then using a 
least-squares fit to obtain the force constants. All total energy and force 
calculations were performed using the Vienna \textit{ab initio} Simulation Package (VASP) 
$^{30, 31}$, which implements Bl\"{o}chl's projector augmented wave (PAW) 
approach $^{32, 33}$ within the local density approximation (LDA). The 
calculations used energy cutoffs of 300-400 eV and $\sim $150-200 unique 
$\mathord{\buildrel{\lower3pt\hbox{$\scriptscriptstyle\rightharpoonup$}}\over 
{k}} $-points per 4 atoms. This resulted in energies converged to $\sim $5 
meV, and forces converged to 1{\%}. Special care was taken to make sure 
forces on the initial structure vanished before supercell configurations 
were generated for the force constant calculation.

The superstructures and their perturbations were determined using an 
efficient algorithm $^{26}$, which generates supercells with perturbed atoms 
and their corresponding set of linear equations. Third-order anharmonic 
terms, which are present when positive and negative displacements are not 
symmetrically equivalent, were eliminated by subtracting forces taken from 
calculations of perturbations of equal magnitude but opposite sign $^{26}$. 
Vibrational entropy calculations were converged to within 0.015 $k_{B}$: this 
value is obtained by calculating the configuration-dependent part of the 
high-temperature limit of the vibrational entropy per atom, $S_{vib} = - k_B 
\int\limits_0^\infty {g(\omega )\ln \omega \,d\omega } $, for increasing 
force constant range until $S_{vib} $changed by less than 0.015 $k_{B}$. 
Typically, this requires $\sim $4-5 neighbors for L1$_{2}$ structures and 2 
neighbors for SQS-8 structures.

The above procedure was applied to calculate vibrational entropies of 
L1$_{2}$-ordered and disordered phases in the Au$_{3}$Cu, Cu$_{3}$Au, 
Au$_{3}$Pd, Pd$_{3}$Au, Cu$_{3}$Pd, Pd$_{3}$Au, and Cu$_{3}$Pt systems. In 
addition, we calculated vibrational entropies of L1$_{2}$-ordered Ag$_{3}$Au 
and Au$_{3}$Ag. The disordered structure was approximated by an 8 atom 
special quasirandom structure (SQS) $^{34}$. These structures have been 
shown to give the best possible approximation to a disordered structure, 
within a given number of sites. In Ni$_{3}$Al, Morgan et. al. tested the 
convergence of vibrational properties with SQS size, and found that a SQS 
with 8 atoms is a good approximation of the disordered state $^{35}$. The 
SQS-8 used in this study has been used successfully in first-principles 
vibrational entropy calculations of disordered Ni$_{3}$Al and Pd$_{3}$V 
$^{12, 13}$.

To define transferable force constant matrices, some approximations are 
necessary. The form of a force constant matrix, and hence the number of 
non-zero force constants, depends on the symmetry of structure. In this 
work, we use a stretching-bending force constant model. In this model, the 
coordinate system of each force constant matrix is transformed, so that the 
$z$-axis is aligned along the segment joining the two atoms in question. Two 
further approximations are necessary to obtain transferable force constants. 
First, bending terms are averaged, so that they are orientation independent. 
Second, off-diagonal terms are constrained to be zero. The resulting force 
constant matrix has only two independent terms - a stretching term $s$, and a 
bending term $b$ - and the form of the matrix can be written

\begin{equation}
\label{eq1}
\left( {{\begin{array}{*{20}c}
 b \hfill & 0 \hfill & 0 \hfill \\
 0 \hfill & b \hfill & 0 \hfill \\
 0 \hfill & 0 \hfill & s \hfill \\
\end{array} }} \right) \quad .
\end{equation}

The stretching term ($s)$ can be used as a qualitative measure of the bond 
strength. However, it is not possible to keep only the stretching terms 
($s)$ in the force constant matrix: doing so can result in large errors (0.2 
$k_{B})$ in the vibrational entropy $^{12, 26}$. We examined two types of 
errors associated with using stretching-bending force constant matrices: 
errors introduced in the force constants and errors introduced in the 
vibrational entropy. For all force constant matrices examined, the rms error 
associated with constraining the bending terms to be equal was 0.052 
eV/A$^{2}$ per bending term. The rms error associated with setting 
off-diagonal terms to zero was 0.009 eV/A$^{2}$ per off-diagonal term. 
Stretching terms contain no errors associated with the stretching-bending 
force constant model: they are directly obtained from the coordinate 
transformation, with no further approximations. 

The errors introduced into the vibrational entropy by using the 
stretching-bending force constant model are shown in Table 1. For 14 out of 
the 16 structures tested, $S_{vib} $ calculated using full-force constant 
matrices and $S_{vib} $ calculated using the stretching-bending force 
constant model agreed to within 0.01 $k_{B}$. For the L1$_{2}$ Au$_{3}$Cu 
structure, the stretching-bending force constant approximation failed when 
using nearest-neighbor force constants; for the SQS-8 Au$_{3}$Cu structure, 
the stretching-bending force constant approximation failed when using both 
nearest-neighbor and all-neighbor force constants (where ``all-neighbor'' is 
defined as the force constant range at which vibrational entropies were 
converged to within 0.015 $k_{B})$. In these structures, using 
stretching-bending force constant matrices resulted in a dynamically 
unstable structure. Tests of each approximation revealed that the negative 
phonon modes in the L1$_{2}$ Au$_{3}$Cu structure arise because the 
off-diagonal terms to set to zero; negative phonon modes in the SQS-8 
Au$_{3}$Cu arise due to both the neglect of the off-diagonal terms and the 
constraint on the bending terms. In previous work, the errors in vibrational 
entropy by using the simplified stretching-bending force constant model have 
been tested for five structures in the Ni$_{3}$Al and Pd$_{3}$V system 
$^{12, 13}$. In those systems, the errors on vibrational entropy for all 
structures were less than 0.01 $k_{B}$/atom.

\section{Results and Discussion}

Calculated vibrational entropies for all structures examined in the Ag-Ag, 
Au-Cu, Au-Pd, Cu-Pt, and Cu-Pd system are listed in Table 1. For the Cu-Au 
system, our results can be compared with previous experimental $^{10, 36}$ 
and theoretical $^{14}$ work. We calculated $S_{vib} \left( {Cu} \right) = $ 
4.87 $k_{B}$ and $S_{vib} \left( {Au} \right) = $2.85 $k_{B}$, which yields 
for the formation entropies$S_{vib}^{form} (L1_2 \;Cu_3 Au) = $ 0.10 
$k_{B}$ and $S_{vib}^{form} (L{\kern 1pt} 1_2 \;Au_3 Cu) = $ 0.11 $k_{B}$. Our 
calculated $S_{vib}^{form} (L1_2 \;Cu_3 Au)$agrees well with previous 
experimental work, which found $S_{vib}^{form} (L1_2 \;Cu_3 Au) = $ 0.07 
$\pm $ 0.03 $k_{B} \quad ^{36}$. Our calculated formation entropies also agree 
well with previous theoretical work, which found $S_{vib}^{form} (L1_2 
\;Cu_3 Au) = $ 0.10 $k_{B}$ and $S_{vib}^{form} (L{\kern 1pt} 1_2 \;Au_3 Cu) 
= $ 0.14 $k_{B} \quad ^{14}$. The entropy change upon disordering in Cu$_{3}$Au 
$\left( {\Delta S_{vib}^{order \to disorder} \left( {Cu_3 Au} \right)} 
\right)$ is calculated to be 0.07 $\pm $ 0.045 $k_{B}$. This value is lower 
than the experimental value of 0.14 $\pm $ 0.05 $k_{B} \quad ^{10}$. However, 
this calculation is in good agreement with previous theoretical work, in 
which $\Delta S_{vib}^{order \to disorder} \left( {Cu{ }_3Au} \right) = $ 
0.08 $k_{B}$ was obtained $^{14}$. For Au$_{3}$Cu, we calculate $\Delta 
S_{vib}^{order \to disorder} \left( {Au{ }_3Cu} \right) = - 0.01\;\pm 
0.03\,k_{B}$, which is lower than the previously calculated $\Delta 
S_{vib}^{order \to disorder} \left( {Au_3 Cu} \right) = $ 0.05 $k_{B}$ 
$^{14}$. 

Calculated full force constant matrices were transformed to the 
stretching-bending force constant model, with the results plotted in Figures 
1-6. The different bond lengths correspond to equilibrium bond lengths in 
various structures. For all pair types, the force constant stiffness for 
both stretching and bending terms decreases with increasing bond length. In 
particular, the stiffness of first nearest-neighbor force-constant matrices 
shows a strong dependence on bond length. This dependence of bond stiffness 
on bond length shows that one can not rely too much upon simple 
bond-counting arguments $^{26}$ when predicting or explaining vibrational 
entropy differences between phases because the bond counting effect does not 
take into account bond-length changes upon disordering. Figures 1-6 show 
that small changes in bond length upon disordering can have a large effect 
on force constants and vibrational entropy. The relationship between atomic 
relaxations and vibrational entropy has been previously noted $^{12, 16 - 
18}$. 

For a given bond type, stretching and bending force constants for all 
systems and structures are found to lie on a single curve. Thus, the 
dependence of bond stiffness on ordering, composition, or chemical system 
can be explained almost entirely in terms of bond length changes. 
Remarkably, other effects on force constant stiffness, such as changes in 
charge density associated with different configurations and chemical 
systems, are small. This may be because our study is limited to noble-metal 
intermetallics.

For each pair type, we constructed a relationship between force constant 
stiffness and bond length. Our goal is to parameterize the stiffness of the 
stretching and bending force constants as a function of bond length, then 
use ``fitted'' force constants from these functions to predict vibrational 
entropy. Stretching terms were fit to a second-order polynomial; bending 
terms were fit to a linear function. The fit was restricted to first-nearest 
neighbors force constants. The rms fitting error over all stretching terms 
was 0.102 eV/A$^{2}$; the rms fitting error over all bending terms was 0.048 
eV/A$^{2}$.

Although attempts to use longer-ranged fitted force constants were 
unsuccessful, the errors introduced by using only nearest-neighbor force 
constants was small. In the systems studied, vibrational entropy converged 
quickly with respect to neighbors: the difference between $S_{vib} $ 
calculated using only first nearest-neighbors and $S_{vib} $ calculated 
using all neighbors was typically around $\sim $0.00-0.02 $k_{B}$,$_{ }$with 
a maximum difference of 0.03 $k_{B}$ in L1$_{2}$ Pd$_{3}$Cu. Previous studies 
on intermetallics and group-IV semiconductors have also shown that 
vibrational entropy can converge quickly with respect to neighbors $^{12, 
13, 37}$. 

The results of using fitted force constants to predict vibrational entropies 
are shown in Table 1. For all structures, the vibrational entropies obtained 
using fitted force constants agree well with the vibrational entropies 
obtained using directly calculated force constants. The rms error between 
$S_{vib}^{calc,all\,neigh,\,full\,fc} $, and $S_{vib}^{fit,1nn,\,sb} $ for 
all structures investigated was 0.032 $k_{B, }$with a maximum error of 0.058 
$k_{B}$ in L1$_{2}$ Pd$_{3}$Au. The errors are $\sim $1-2{\%} of the absolute 
vibrational entropy. These calculations show the predictive power of using 
fitted stretching-bending force constants to predict vibrational entropies. 

We also tested the accuracy of using fitted force constants, in representing 
the effect of homogeneous volume changes. In metals, pairwise expansion can 
usually not capture the effect of such electron density changes of the 
energy. The force constants and vibrational entropy of L1$_{2}$ and SQS-8 
structures for Pd$_{3}$Cu, Pd$_{3}$Au, Cu$_{3}$Pd, Cu$_{3}$Au, and 
Au$_{3}$Pd were recalculated at a volume $\sim $2{\%} larger than the 
equilibrium volume. For the L1$_{2}$ Pd$_{3}$Cu structure, the vibrational 
entropies were also calculated at a 4{\%} and 6{\%} volume increase were 
also calculated. These direct results were then compared with predicted 
results obtained from using force constants derived from equilibrium volume 
calculations. None of the force constants at these increased volumes were 
included in the fit. The results of using these fitted force constants to 
predict vibrational entropy are shown in Table 2. For all structures at 
2{\%} increased volume, the rms error between 
$S_{vib}^{calc,all\,neigh,\,full\,fc} $ and $S_{vib}^{fit,1nn,\,sb} $ was 
0.033 $k_{B}$, with a maximum error of 0.053 $k_{B}$ in L1$_{2}$ Pd$_{3}$Au. 
Thus, even when the force constant data is not included in the fit, the 
length dependent force constant function gives force constants that 
accurately predict vibrational entropies. 

Often, one is interested in vibrational entropy differences, rather than 
absolute vibrational entropies. We used vibrational entropy data from Tables 
1-2 to obtain two types of vibrational entropy differences that are of 
interest: the vibrational entropy difference between ordered and disordered 
phases, and vibrational entropy difference between a structure at two 
volumes separated by $\sim $2{\%}. The latter is related to the thermal 
expansion by the expression $\alpha _L = \frac{\Delta S}{3B\Delta V}$, where 
B is the bulk modulus. This data is shown in Table 3. In the systems 
studied, we estimated the effect of vibrations on calculated phase 
boundaries by using the following procedure: The change in transition 
temperature when vibrations are included in the phase diagram calculation is 
given by $T_{config + vib}^{\alpha \to \beta } = T_{config\,only}^{\alpha 
\to \beta } \left( {1 + \frac{\Delta S_{vib}^{\alpha \to \beta } }{\Delta 
S_{config}^{\alpha \to \beta } }} \right)^{14}$. Thus, $\frac{\Delta 
S_{vib}^{\alpha \to \beta } }{\Delta S_{config}^{\alpha \to \beta } }$ 
determines the effect of lattice vibrations on calculated phase diagrams. 
The configurational entropy per atom for a binary solid solution depends on 
the state of short-range order. It has a maximum value for an ideal (random) 
solution of $S_{config}^\alpha = k\left[ {c\ln c + \left( {1 - c} \right)\ln 
(1 - c)} \right]$. Hence, the maximum value of $\Delta S_{config}^{\alpha 
\to \beta } $ in the systems studied is 0.562 $k_{B}$/atom: this value occurs 
when a fully ordered state ($S_{config}^\alpha = 0)$ at $c$ = 0.25 transforms 
to a fully random solid solution. In most cases, the actual value of $\Delta 
S_{config}^{\alpha \to \beta } $ will be smaller than this, owing to 
short-range order in the disordered state and also some disorder in the low 
temperature phase. Nevertheless, our approximate estimate shows that phase 
boundaries in the systems studied would change by $\sim $5-13{\%} if 
vibrations were included. 

The data in Table 3 can be used to examine the ability of transferable force 
constants to predict vibrational entropy differences. The rms error between 
calculated and fit differences $\left( {\Delta S_{vib}^{{\kern 1pt} 
calc,\,all\,neigh,\,full\,fc} - \Delta S_{vib}^{fit,\,1nn,\,sb} } \right)$ 
was 0.030 $k_{B,}$ with a maximum error of 0.063 $k_{B}$ in Cu$_{3}$Au. For 
vibrational entropy differences between structures at volumes 2{\%} apart, 
the rms error between calculated and fit differences was 0.016 $k_{B}$ with a 
maximum error of 0.030 $k_{B}$ in L1$_{2}$ Au$_{3}$Pd. Thus, the transferable 
force constants used here are able to predict vibrational entropy 
differences reasonably well.

In examining the data, one can make a few observations with respect to the 
accuracy of using transferable force constants. First, errors in absolute 
entropies from using transferable force constants (Tables 1-2) were 
typically $\sim $0.00-0.05 $k_{B }$for all structures. This was true for 
structures with both large and small absolute vibrational entropies. Second, 
in all cases, the transferable force constants were able to correctly 
predict the relative hierarchy of entropy differences. Thus, large entropy 
differences are predicted to be large; small entropy differences are 
predicted to be small. A small entropy difference was never predicted to be 
large, or vice versa. Third, when using transferable force constants, errors 
in entropy differences between the same structure at different volumes 
tended to cancel; errors in entropy differences between different structures 
(at the same composition) did not tend to cancel as much (Table 3). Thus, it 
is likely that there are small structure-dependent contributions to the 
force constant stiffness that are not captured by our length-dependence 
force constant model. Fourth, using transferable force constants gave small 
absolute errors on vibrational entropy differences ($\sim $0.00-0.07 
$k_{B})$. For small vibrational entropy differences, this sometimes resulted 
in $\sim $15-20{\%} errors, whereas the percentage errors in larger 
vibrational entropy differences were smaller. Lastly, this method can 
greatly reduce the computational cost of calculating vibrational properties 
of intermetallics.

\section{Conclusion}

The vibrational entropies for 16 structures in the Ag-Au, Au-Cu, Au-Pd, 
Cu-Pd, and Cu-Pt systems have been calculated. A simplified model was used 
with only stretching and bending terms, making the force constant matrices 
independent of symmetry. With these approximations, the form of the force 
constant matrices was independent of crystal symmetry. We found that the 
variation of force constants with ordering, composition, or chemical system 
can be explained almost entirely through changes in bond length. This method 
represents a promising way to include vibrational effects in phase diagram 
calculations at a moderate computational cost.

\section*{Acknowledgements}

This work was supported by Department of Energy, Office of Basic Energy 
Sciences under Contract No. DE-FG02-96ER45571. We gratefully acknowledge 
computing resources provided by NPACI through the Texas Advanced Computing 
Center. We would also like to thank postdoctoral researcher Dane Morgan for 
insightful observations.

$^{1 }$J. M. Sanchez, F. Ducastelle, and D. Gratias, Physica A, 
\textbf{128a}, 334 (1984).

$^{2 }$F. Ducastelle, \textit{Order and phase stability in alloys}, Elsevier Science Pub. Co., Amsterdam, 1991.

$^{3 }$D. de Fontaine, Solid State Physics, \textbf{47}, 33 (1994).

$^{4 }$G. Ceder, M. Asta, and D. de Fontaine, Physica C, \textbf{177}, 106 
(1991).

$^{5 }$A. Zunger, in \textit{NATO ASI on Statics and Dynamics of Alloy Phase Transformations},edited by P. E. Turchi and A. Gonis Plenum Press, New 
York, 1994, Vol. p.361.

$^{6 }$M. Asta, R. McCormack, and D. de Fontaine, Phys. Rev. B, \textbf{48}, 
748 (1993).

$^{7 }$B. Fultz, L. Anthony, L. J. Nagel, et al., Phys. Rev. B, \textbf{52}, 
3315 (1995).

$^{8 }$L. Anthony, J. K. Okamoto, and B. Fultz, Phys. Rev. Lett., 
\textbf{70}, 1128 (1993).

$^{9 }$L. Anthony, L. J. Nagel, J. K. Okamoto, et al., Phys. Rev. Lett., 
\textbf{73}, 3034 (1994).

$^{10 }$L. J. Nagel, L. Anthony, and B. Fultz, Philos. Mag. Lett., 
\textbf{72}, 421 (1995).

$^{11 }$M. E. Manley and B. Fultz, Philos. Mag. B, \textbf{80}, 1167 (2000).

$^{12 }$A. van de Walle and G. Ceder, Phys. Rev. B, \textbf{61}, 5972 
(2000).

$^{13 }$A. van de Walle, G. Ceder, and U. V. Waghmare, Phys. Rev. Lett., 
\textbf{80}, 4911 (1998).

$^{14 }$V. Ozolins, C. Wolverton, and A. Zunger, Phys. Rev. B, \textbf{58}, 
R5897 (1998).

$^{15 }$P. D. Tepesch, A. F. Kohan, G. D. Garbulsky, et al., J. Am. Ceram. 
Soc., \textbf{79}, 2033 (1996).

$^{16 }$V. Ozolins and M. Asta, Phys. Rev. Lett., \textbf{86}, 448 (2001).

$^{17 }$M. Asta and V. Ozolins, Phys. Rev. B, \textbf{64}, art. no. 094104 
(2001).

$^{18 }$C. Wolverton and V. Ozolins, Phys. Rev. Lett., \textbf{86}, 5518 
(2001).

$^{19 }$G. D. Garbulsky and G. Ceder, Phys. Rev. B, \textbf{49}, 6327 
(1994).

$^{20 }$I. M. Robertson, J. Phys.: Condens. Matter , \textbf{3}, 8181 
(1991).

$^{21 }$M. Sluiter, M. Weinart, and Y. Kawazoe, Europhys. Lett., 
\textbf{43}, 183 (1998).

$^{22 }$M. Sluiter and Y. Kawazoe, Philos. Mag. A, \textbf{78}, 1353 (1998).

$^{23 }$M. H. F. Sluiter, M. Weinert, and Y. Kawazoe, Phys. Rev. B, 
\textbf{59}, 4100 (1999).

$^{24 }$P. Ghosez, E. Cockayne, U. V. Waghmare, et al., Phys. Rev. B, 
\textbf{60}, 836 (1999).

$^{25 }$P. Giannozzi, S. de Gironcoli, P. Pavone, et al., Phys. Rev. B, 
\textbf{43}, 7231 (1991).

$^{26 }$A. van de Walle and G. Ceder, Rev. Modern Phys., \textbf{74}, 11 
(2002).

$^{27 }$S. Wei and M. Y. Chou, Phys. Rev. Lett., \textbf{69}, 2799 (1992).

$^{28 }$S. Wei and M. Y. Chou, Phys. Rev. B, \textbf{50}, 2221 (1994).

$^{29 }$K. Kunc and R. M. Martin, Phys. Rev. Lett., \textbf{48}, 406 (1982).

$^{30 }$G. Kresse and J. Furthmuller, Comp. Mater. Sci., \textbf{6}, 15 
(1996).

$^{31 }$G. Kresse and J. Furthmuller, Phys. Rev. B, \textbf{54}, 11169 
(1996).

$^{32 }$G. Kresse and D. Joubert, Phys. Rev. B, \textbf{59}, 1758 (1999).

$^{33 }$P. E. Blochl, Phys. Rev. B, \textbf{50}, 17953 (1994).

$^{34 }$A. Zunger, S. H. Wei, L. G. Ferreira, et al., Phys. Rev. Lett., 
\textbf{65}, 353 (1990).

$^{35 }$D. Morgan, J. D. Althoff, and D. de Fontaine, J. Phase Equilib., 
\textbf{19}, 559 (1998).

$^{36 }$P. D. Bogdanoff and B. Fultz, Philos. Mag. B, \textbf{79}, 753 
(1999).

$^{37 }$G. Garbulsky, Ph.D. Thesis, http://africa.mit.edu/papers.

\newpage

\illuseps{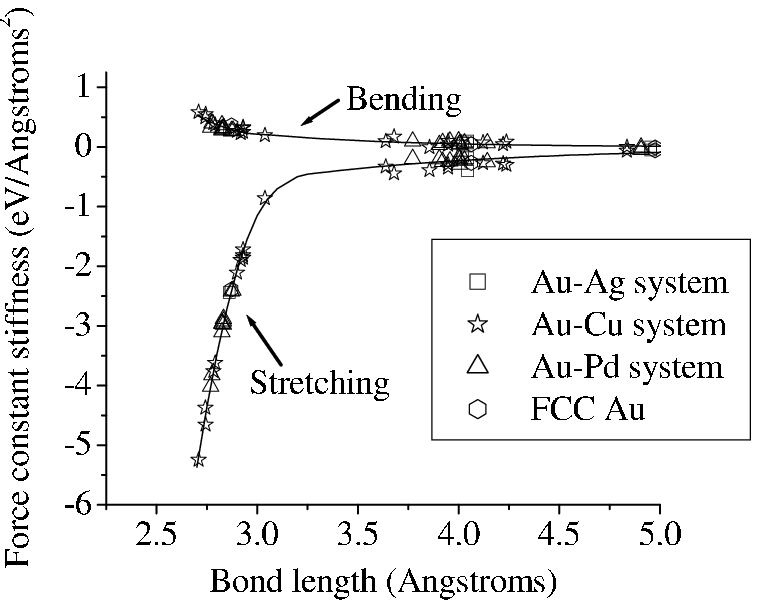}{Stiffness of bending and stretching force
constants vs. bond length for Au-Au bonds. Force constants for the
same system are represented by the same symbol (for example, all Au-Au
bonds for L1$_{2}$ Au$_{3}$Cu use the same symbol). Lines are drawn as
guides to the eye.}

\illuseps{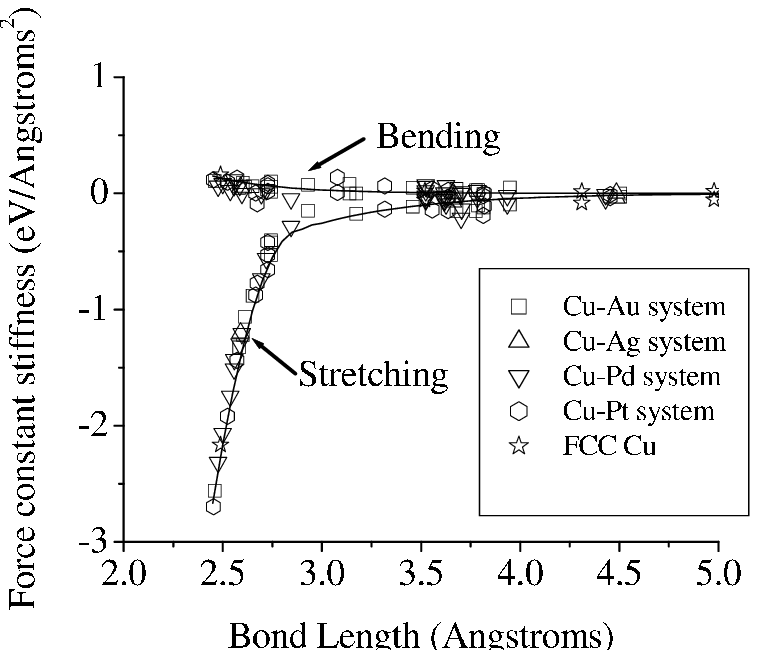}{Stiffness of bending and stretching force constants vs. bond 
length for Cu-Cu bonds. Lines are drawn as guides to the eye.}

\illuseps{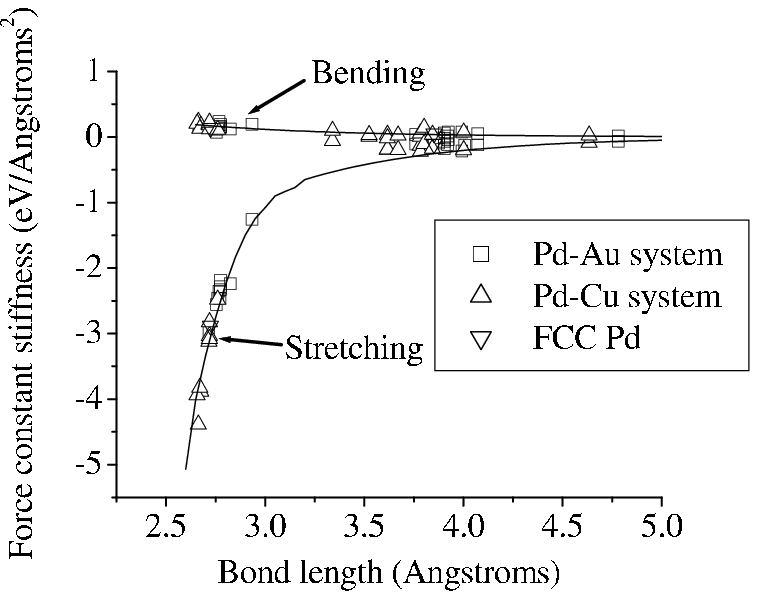}{Stiffness of bending and stretching force constants vs. bond 
length for Pd-Pd bonds. Lines are drawn as guides to the eye.}

\illuseps{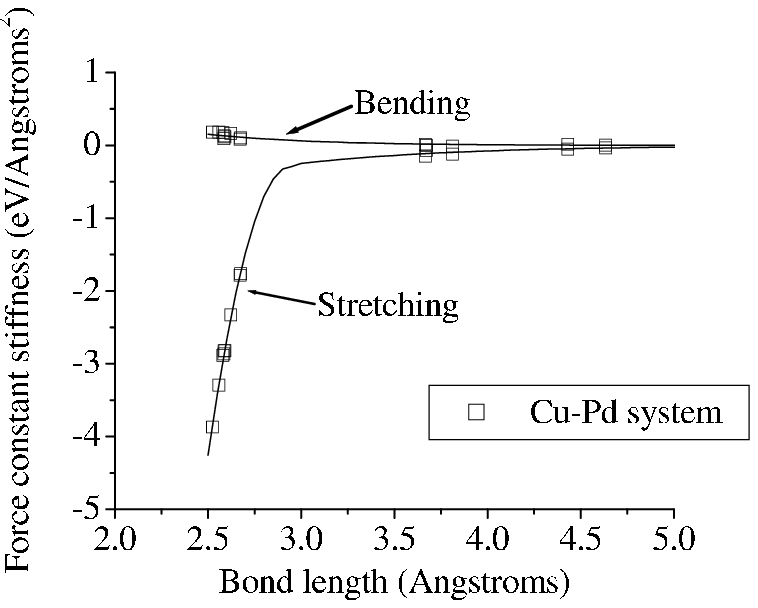}{Stiffness of bending and stretching force constants vs. bond 
length for Cu-Pd bonds. Lines are drawn as guides to the eye.}

\illuseps{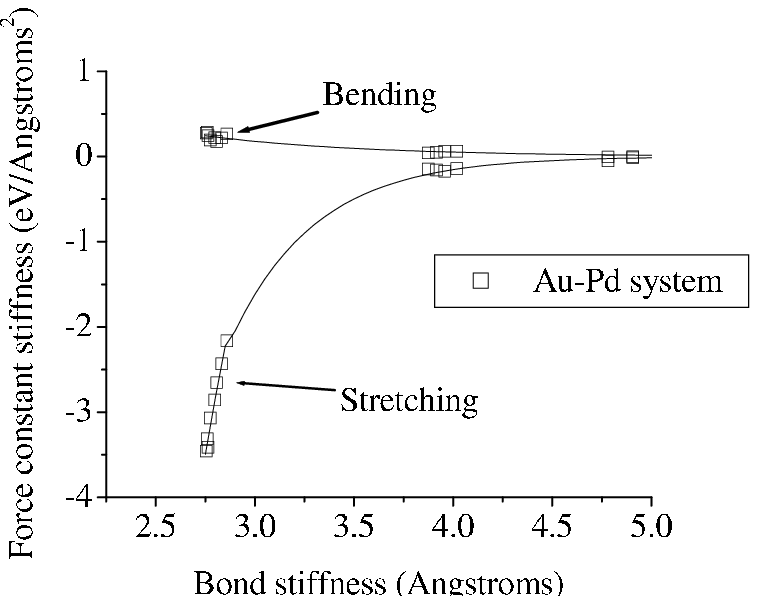}{Stiffness of bending and stretching force constants vs. bond 
length for Au-Pd bonds. Lines are drawn as guides to the eye.}

\illuseps{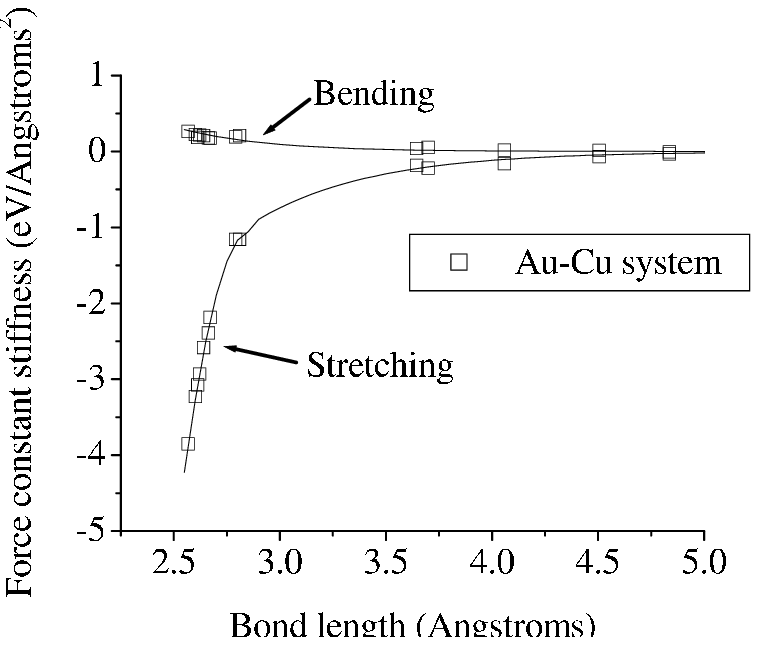}{Stiffness of bending and stretching force constants vs. bond 
length for Au-Cu bonds. Lines are drawn as guides to the eye.}

\newpage 

\tableeps{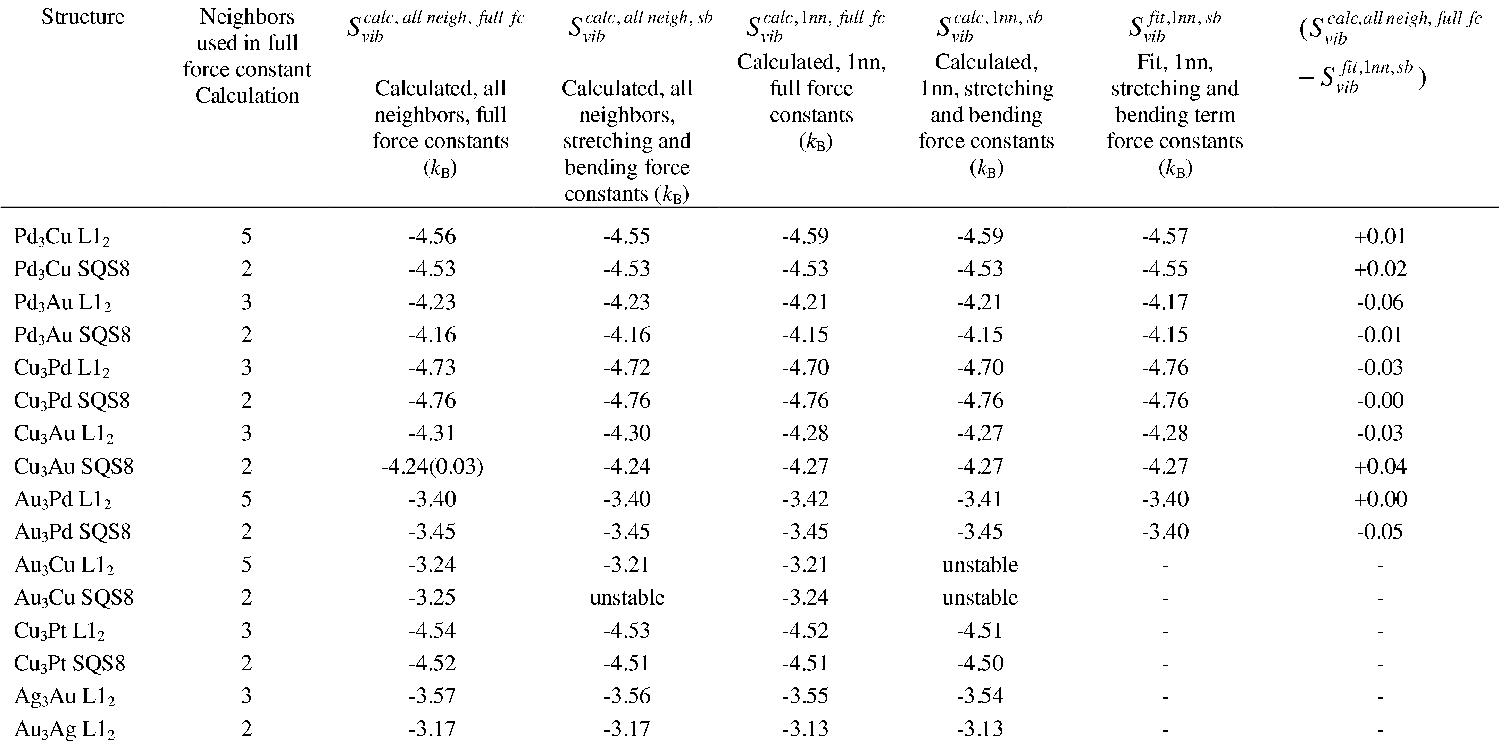}{Vibrational entropy data. Listed are vibrational entropies 
calculated with all neighbors and full force constant matrices, vibrational 
entropies calculated with all neighbors and stretching-bending force 
constants, calculated vibrational entropy with first-nearest neighbors and 
full force constant matrices, calculated vibrational entropy with 
first-neighbors and stretching-bending force constant matrices, and 
fit-vibrational entropies using first-neighbors and stretching-bending force 
constants. Also listed is$(S_{vib}^{calc,all} - S_{vib}^{fit,1nn} )$, the 
error introduced by using fit force constants, the stretching-bending force 
constant model, and first-nearest neighbor force constants. Errors are 0.015 
$k_{B}$ unless otherwise indicated. All numbers are rounded to 0.01 
$k_{B}$.}

\tableeps{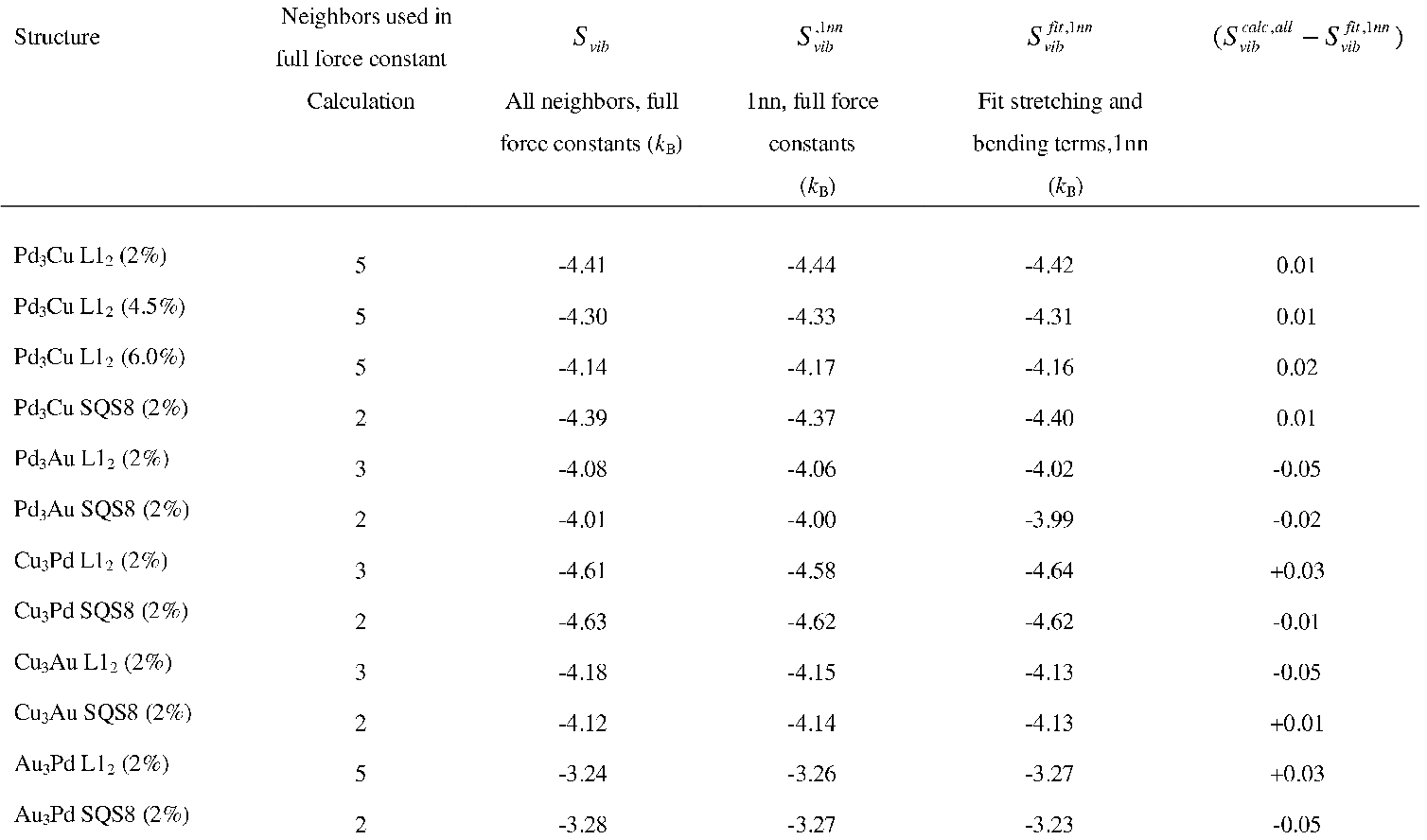}{Vibrational entropies at increased
volume. Percentage increase in volume is indicated in
parenthesis. Listed are vibrational entropies calculated with all
neighbors and full force constant matrices, vibrational entropies
calculated with fit first-nearest neighbor stretching-bending force
constants, and $(S_{vib}^{calc,all\,neigh,\,full\,fc} -
S_{vib}^{fit,1nn,\,sb} )$. All numbers are rounded to 0.01 $k_{B}$.}

\tableeps{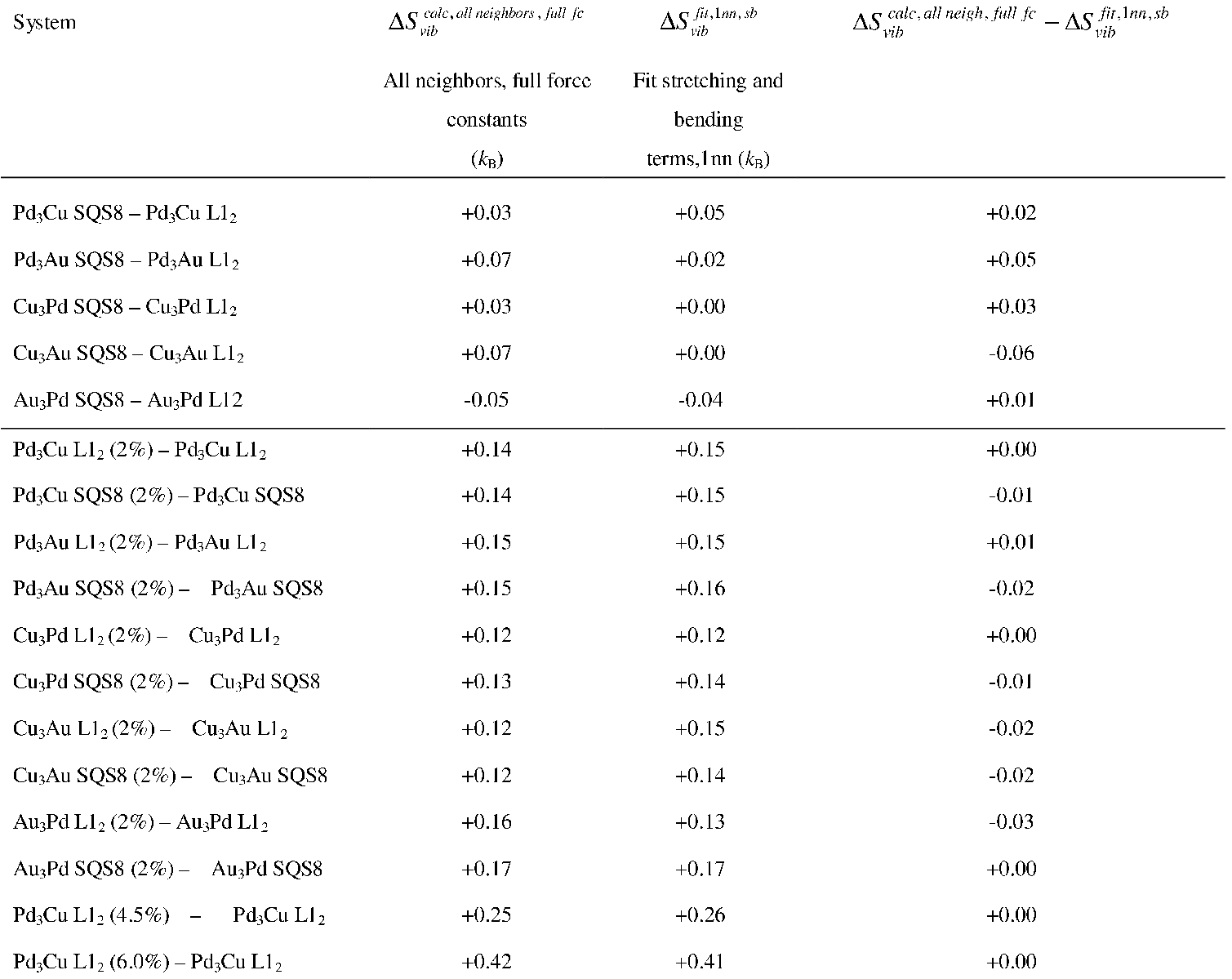}{Vibrational entropy differences. Listed are
vibrational entropy differences upon disordering and vibrational
entropy differences between the same structure at different
volumes. Percentage increase in volume is indicated in
parenthesis. Also shown are errors in differences, $\Delta
S_{vib}^{{\kern 1pt} calc,\,all\,neighbors} - \Delta
S_{vib}^{fit,\,1nn} $.}

\end{document}